\documentstyle[12pt,aps,prd]{revtex}
\tightenlines
\input epsf

\newcommand{\beq}{\begin{equation}}
\newcommand{\eeq}{\end{equation}}

\def\lap{\lower.5ex\hbox{$\; \buildrel < \over \sim \;$}}
\def\gap{\lower.5ex\hbox{$\; \buildrel > \over \sim \;$}}
\def\lesssim{\lap}

\begin{document}

\title{Cosmic strings: progress and problems}
\author{Alexander Vilenkin}
\address{
Institute of Cosmology, Department of Physics and Astronomy,\\
Tufts University, Medford, MA 02155, USA\\
}

\maketitle

\begin{abstract}

Recent developments in cosmic strings are reviewed, with emphasis on
unresolved problems.

\end{abstract}

\section{Introduction}

It gives me great pleasure to contribute to the volume honoring
Katsuhiko Sato at his 60th birthday. It is a bit embarrassing that the
subject of my contribution is one of the very few topics in modern
cosmology (and probably the only one in these Proceedings) on which
Katsu has not worked at one time or another during his career. But
then this only serves to illustrate the amazingly wide range of his
research. 

The idea of cosmic strings was popular in the 1980's and early 90's,
mainly in the context of structure formation theory. Density
perturbations that would be produced by strings of the grand
unification energy scale would have the right magnitude to serve as
seeds for galaxies and clusters. This scenario was then disfavored by
CMB observations, and cosmic string research was all but abandoned for
nearly a decade. Recently, however, there has been a revival of
interest in the subject. This string Renaissance has been mostly
fueled by the developments in superstring theory. It now appears that
fundamental strings may have astronomical dimensions and play the role
of cosmic strings. Also, it has been realized that cosmic strings can
have observable effects even if their energy scale is well below grand
unification. Furthermore, there are some intriguing indications that a
cosmic string may have already been observed. Here, I will briefly
review the present state of the art, emphasizing recent developments
and unresolved problems.

\section{``Ordinary'' strings}

Cosmic strings can be formed as linear defects at symmetry breaking
phase transitions in the early universe. Their formation and evolution
have been extensively discussed in the literature; for a review and
references see \cite{VS,HK}.

Topologically stable strings do not have ends. They can either form
closed loops or extend to infinity. The mass per unit length of
string, $\mu$, which is also equal to the string tension, is of the
order $\mu\sim\eta^2$, where $\eta$ is the energy scale of symmetry
breaking. The dimensionless parameter $G\mu\sim \eta^2/M_p^2$, where
$G$ is Newton's constant and $M_p$ is the Planck mass, characterizes
the strength of gravitational interactions of strings. For
grand-unification-scale strings, $G\mu\sim 10^{-6}$.

At the time of formation, strings have the form of a tangled network,
consisting of Brownian infinite strings and a distribution of closed
loops. Curved strings move under the action of their tension,
developing speeds close to the speed of light. When strings cross,
they reconnect, or ``exchange partners''. Self-intersections result 
in the formation of closed loops. Numerical simulations
of string evolution indicate that the network of long strings evolves
in a scale-invariant fascion. The typical distance between the strings
$d(t)$, and the coherence length $\xi(t)$, defined as the distance
beyond which the directions along the string are uncorrelated, both
scale with the cosmic horizon:
\begin{equation}
d(t)\sim\xi(t)\sim t.
\label{scaling}
\end{equation}

Long strings exhibit significant small-scale structure, which is
partly a remnant of the initial wiggliness of the strings and partly
due to multiple kinks formed at string intersections. As strings move
at relativistic speeds, each horizon-size segment intersect
itself or another long string about once in a Hubble time $t$. As a
result, one or few loops of size $\sim t$ are produced per Hubble
volume per Hubble time. These large loops then shatter, through
multiple self-intersections, into a large number of small
loops. According to the standard lore, the characteristic size of
loops, $l_{loop}(t)$ is set by the typical wavelength of the smallest
wiggles on long strings, $l_{wiggles}(t)$. There is, however, no hard
evidence for that.  In string simulations performed in the 1990's,
both $l_{wiggles}$ and $l_{loop}$ remained at the resolution of the
simulations, so neither of these scales could be reliably determined.

There have been claims that long strings lose a substantial amount of
energy by direct emission of tiny loops of size not much greater than
the string thickness \cite{Vincent} and by radiation of massive
particles \cite{Hindmarsh}. Personally, I find this hard to
believe. It is not clear how the motion of astronomically large
strings can excite the extremely short-wavelength perturbations on the
scale of the string thickness. This view is supported by the numerical
results in \cite{More,Jose}, which show no evidence for massive
radiation.

The standard scenario of string evolution assumes that the main
mechanism of string smoothing on the smallest scales is the
gravitational damping of small-scale wiggles. A simple estimate then
gives 
\begin{equation}
l_{wiggles}(t)\sim l_{loop}(t)\sim \alpha t,
\label{wiggles}
\end{equation}
where 
\begin{equation}
\alpha\sim 50G\mu.
\label{alpha}
\end{equation}

Recent developments, however, put into question some of the key
assumptions of the standard scenario. First, it has been realized that
the gravitational radiation from counter-streaming wiggles on long
strings is far less efficient at damping the wiggles than originally
thought \cite{OS}. If indeed $\alpha$ is determined by the
gravitational back-reaction, then the new analysis shows \cite{OSV}
that its value is sensitive to the spectrum of small-scale wiggles and
is generally much smaller than (\ref{alpha}). Moreover,
high-resolution numerical simulations of strings in flat spacetime
indicate that small wiggles evolve in a scale-invariant manner,
even in the absence of gravitational damping \cite{VOV}. This suggests
that the parameter $\alpha$ may be unrelated to $G\mu$, at least for
strings of very low energy scale. The time is now ripe for a new
generation of string simulations, which may help to resolve these
important issues.

\section{Cosmic superstrings}

Witten \cite{witten} was the first to consider the possibility of
cosmic superstrings, but only to promptly rule it out. For fundamental
strings, the mass per unit length is $\mu\sim M_s^2$, where $M_s$ is
the string energy scale. If one assumes that $M_s\sim M_p$, then $G\mu\sim
(M_s/M_p)^2 \sim 1$. This is far above the observational bounds, which
require that $G\mu\lesssim 10^{-7}$.

Recently, however, it has been shown \cite{AHDD} that in models with
large extra dimensions $M_s$ can be $\ll M_p$, yielding $G\mu\ll
1$. Another possibility is that the effective string tension may be
small due to the bulk gravitational potentials, or ``warp factors'',
as in the Randall-Sundrum model \cite{Randall}. 
Suppose the metric is of the form
\begin{equation}
ds^2 = F(y)\eta_{\mu\nu}dx^\mu dx^\nu + ds_y^2,
\end{equation}
where $x^\mu$ are the $4D$ spacetime coordinates and $y$ labels the coordinates
in the extra dimensions. If the strings are localized in a
gravitational potential well at $y=y_0$, then their effective tension
is $\mu_{eff}=F(y_0)\mu$. With $F(y_0)\ll 1$, we can have $G\mu_{eff}
\ll 1$ even for $M_s\sim M_p$.

Apart from the fundamental, or $F$-strings, superstring theory
provides another candidate for the role of cosmic strings. It is a
$D$-string, which can either be a 1-dimensional $D$-brane, or a
higher-dimensional brane with all but one dimensions compactified.

Both $F$ and $D$-strings are naturally formed in models of brane
inflation. In such models, the inflationary expansion is driven by the
attractive interaction potential between parallel $D$-brane and
anti-$D$-brane, which are separated in extra dimensions. The two
branes are slowly pulled toward one another, until they collide and
annihilate \cite{DT}. The role of the inflaton field in this model is
played by the separation between the branes. This field becomes
tachyonic when the branes get close together, and the branes quickly
annihilate.

Each brane has a $U(1)$ gauge field living on its worldsheet, so there
is a $U(1)\times U(1)$ symmetry prior to brane annihilation. The
tachyon field is coupled to the combination $(A_1 - A_2)$ of the two
gauge fields, and Nielsen-Olesen vortices form as in the usual Higgs
model when the tachyon develops an expectation value. These vortices
are identified with $D$-strings. The orthogonal combination $(A_1 +
A_2)$ is not higgsed. When the branes annihilate, the electric
component of this field combination is squeesed into electric flux
lines, which are identified with $F$-strings. A substantial fraction
of energy of the annihilating branes can go into a stochastic string
network. However, the details of the string formation mechanism are
not yet fully understood and some points remain controversial
\cite{Tye,DV,CMP}. For a more detailed review of cosmic superstrings,
see \cite{Polchinski}.

Cosmic $F$ and $D$-strings are exciting new alternatives, but it
should be emphasized that we have no reason to dismiss the good old
field-theory strings resulting from symmetry breaking. It has been
argued in \cite{Jeannerot} that such strings inevitably arise in a
wide class of supersymmetric grand unified models which exhibit
inflation.

\section{String properties}

The values of $G\mu$ for $F$- or $D$-strings resulting from brane
inflation are expected to be in the range \cite{JST,Polchinski}
\begin{equation}
10^{-11}\lesssim G\mu\lesssim 10^{-6}.
\label{Gmu}
\end{equation}
This estimate was obtained assuming that the same period of brane 
inflation accounts also for the observed density inhomogeneities. 
$G\mu$ can be much smaller if the perturbations were generated by 
some other mechanism.

An important difference between fundamental and ``ordinary'' strings
is in their reconnection properties. As I already mentioned, when
ordinary strings cross, they always reconnect. For $F$-strings, on the
other hand, reconnection is a quantum-mechanical process, whose
probability is governed by the string tension $g_s$, $P\sim
g_s^2$. Another difference is due to the higher-dimensional habitat of
$F$-strings. Moving linear objects generically intersect in 3
dimensions, but can easily miss one another in higher dimensions. This
leads to further suppresion of reconnection probability \cite{DV,JST}.
A detailed analysis in \cite{JJP} suggests that $10^{-3}\lesssim P
\lesssim 1$ for $F$-strings and $0.1\lesssim P\lesssim 1$ for
$D$-strings.

An intersection of an $F$-string with a $D$-string cannot result in
reconnection. Instead, segments of the two strings can join, forming a
bound state - an $FD$-string. Such interactions between $F$ and $D$
networks can result in the formation of an interconnected web - an
$FD$-network - with $F$, $D$, and $FD$-strings joining at three-way
vertices \cite{CMP,DV}.

In braneworld models, it is also important to consider the interaction
of strings with $D$-branes. $F$-strings can end on a $D$-brane, with
their endpoints playing the role of particles on the brane worldsheet.
Even though brane-antibrane pairs may have annihilated at the end of
inflation, some branes must have survived, since the Standard Model
particles presumably live on a stack of branes. If Brownian
$F$-strings are allowed to wiggle around in the bulk, they will have
multiple intersections with the surviving branes and will quickly
break up into pieces with their ends attached to the branes. The
segments will oscillate under the action of the string tension and
will dissipate their energy by radiation of gauge quanta,
gravitational waves, and other light fields.

This quick demise of an $F$-string network is not inevitable. The
network may be localized away from the surviving branes by bulk
gravitational fields. For example, the brane annihilation, which is
followed by string formation, may occur in one potential well, while
the surviving branes may reside in another well. This is precisely
what happens in the KKLMMT model of brane inflation \cite{KKLMMT}. In
this case, $F$-strings can break up only if a piece of string tunnels
through the potential barrier separating the two wells. The tunneling
may be strongly suppressed, in which case the strings are practically
stable \cite{CMP}.

Depending on the model, $D$-strings may or may not be able to break on
surviving branes. If they can break, then the situation is identical
to that of $F$-strings, and a stable network can exist only if the
strings are separated from the branes by a potential barrier. An
interesting alternative is the case when the bulk potentials localize
$D$-strings on top of the branes. If the strings cannot break (as,
e.g., in the case of $D1$-strings and $D7$-branes \cite{Polchinski}),
they will be superconducting, with open $F$-strings connecting $D1$
and $D7$ playing the role of massless charge carriers \cite{DV}.

\section{String evolution}

The evolution of cosmic superstrings is rather model-dependent. One
possibility is that the strings break up on the surviving branes and
completely disappear before present. 

A more interesting alternative is when strings of only one type
-either $F$ or $D$ - survive until present. The strings would then
evolve as ``ordinary'' cosmic strings, except the reconnection
probability will generally be less than one and can even be $P\ll 1$.  
For a low reconnection probability, the number of long strings per horizon
is expected to increase. The scaling regime of evolution can be
sustained only if the long strings have one or few reconnections per
Hubble distance per Hubble time. With $P\ll 1$, it will take many
string crossings to get a reconnection. Hence, the number of strings
per horizon should be large, $N_s\gg 1$. Simple estimates suggest
\cite{Damour2,Maria} $N_s\sim P^{-1/2}$, in agreement with earlier
numerical simulations \cite{SaVi}. This feature of superstrings, that
there may be many of them in our Hubble volume, may help to
distinguish them observationally from ordinary strings.

If both $F$- and $D$-strings are formed and are confined to the same
potential well in extra dimensions, the strings will combine to form
an $FD$-network. Analytic and numerical models indicate that such
interconnected networks evolve towards a scaling regime, where the
typical inter-string distance remains a fixed fraction of the horizon
\cite{VV,SP,CS,TWW},
\begin{equation}
d(t)\sim\zeta t.
\label{zeta}
\end{equation}
The energy density of the network is $\rho_{FD}\sim
\mu/d^2(t)$. Requiring that it is much smaller than the total energy
density of the universe yields the condition
\begin{equation}
\rho_{FD}/\rho\sim G\mu /\zeta^2\ll 1.
\label{rhoFD}
\end{equation}

The parameter $\zeta$ in (\ref{zeta}) is determined by the rate of
energy dissipation in the network \cite{VV}. The simulations of
Refs. \cite{SP,CS} were performed for global strings,
which dissipate energy very efficiently, through the emission of
Goldstone boson radiation. The corresponding values of $\zeta$ are
$\zeta\sim 0.1-0.01$, and the condition (\ref{rhoFD}) is satisfied for
reasonable values of $G\mu$. For superstring networks, on the other
hand, the dominant radiation mechanism is gravitational radiation. It
has been argued in \cite{VV} that in this case $\zeta\sim G\mu$, and
Eq. (\ref{rhoFD}) gives $\rho_{FD}/\rho\sim 1/G\mu\gg 1$, indicating that
the string network becomes so dense that it dominates the universe.

The conclusion appears to be that models predicting the formation of
$FD$-networks are ruled out. The only caveat is that the networks may
lose energy more efficiently through chopping off small nets, in a way
similar to chopping off closed loops by ordinary strings. This issue
needs to be further investigated in high-resolution simulations of
evolving string networks.

\section{Looking for cosmic strings}

Apart from the new insights from superstring theory, there is another,
independent reason for the renewed interest in cosmic strings. As I
already mentioned, it has been recently realized that gravitational
waves from strings may be detectable for a very wide range of $G\mu$,
extending to values well below the grand unification scale. This is
particularly interesting now, when LIGO and other detectors are beginning
their search for gravitational waves (GW).

GW from string loops oscillating at different cosmic
epochs add up to a stochastic background, which
spans many orders of magnitude in frequency. Such a background would
introduce noise into the arrival times of signals coming from the
millisecond pulsar. The bound on $G\mu$ from the pulsar
data is \cite{Kaspi} 
\begin{equation}
G\mu\lesssim 10^{-7}.
\label{bound}
\end{equation}
(For a detailed discussion of the current pulsar bounds, see
\cite{Damour2}.)  
A similar bound follows from the microwave background observations
\cite{Levon,Fraisse}. 

The key new development is the realization that, in addition to a
featureless Gaussian component, the GW background from strings
includes sharp GW bursts. Oscillating loops of string typically
develop ``cusps'' once or few times during the period of oscillation,
with the string velocity momentarily reaching the speed of light at
the tip of the cusp. Short GW bursts emanating from cusps carry away
about the same power as the radiation at low frequencies, comparable
to the oscillation frequency of the loop. The analysis in
\cite{Damour} suggests that the bursts should be detectable by LIGO
and LISA for values of $G\mu$ as low as $10^{-12}-10^{-14}$. The
effects of the low reconnection probability, $P\ll 1$, and of the
uncertainty in the loop size parameter $\alpha$ have
been investigated in \cite{Damour2}, with the conclusion that the
predictions of \cite{Damour} are quite robust, at least when the loop
sizes are not suppressed by many orders of magnitude relative to the
standard scenario.

Another intriguing new development is the observation of two nearly
identical galaxies at redshift $z=0.46$ with angular separation of
1.9 arc seconds \cite{Sazhin}. The spectra of the two galaxies
coincide at 99.9\% confidence level \cite{Sazhin2}. The most plausible
interpretation of the data appears to be lensing by a cosmic string
with $G\mu\sim 4\times 10^{-7}$. This estimate assumes a slowly moving
string orthogonal to the line of sight at a relatively low redshift
$(z\lesssim 0.1)$. Increasing the string redshift or changing its
orientation would give a higher estimate for $G\mu$, which might be in
conflict with the CMB and pulsar observations. On the other hand, the
estimate for $G\mu$ could be decreased due to relativistic motion
\cite{AV86,Tye2} or wiggliness \cite{Vach} of the string. The only
alternative to the string interpretation is that two very similar
galaxies just happen to be next to one another. The issue is likely to
be resolved by Space Telescope observations later this year.

If strings are superconducting, they can produce a rich variety of
observational effects. Among them are gamma-ray bursts \cite{Bohdan}
and high-energy cosmic rays (for a discussion and references, see
\cite{Blasi}). 

I am grateful to Thibault Damour and Nick Jones for useful comments. 
This work was supported in part by the National Science Foundation.


\begin{thebibliography}{99}

\bibitem{VS}
A. Vilenkin and E.P.S. Shellard, {\it Cosmic Strings and Other
Topological Defects}, Cambridge University Press (Cambridge, 2000). 

\bibitem{HK}
M.B. Hindmarsh and T.W.B. Kibble, Rep. Prog. Phys. {\bf 58}, 477
(1995). 

\bibitem{Vincent}
G.R. Vincent, M. Hindmarsh and M. Sakellariadou, Phys. Rev. {\bf D56},
637 (1997).

\bibitem{Hindmarsh}
G.R. Vincent, N.D. Antunes and M. Hindmarsh, Phys. Rev. Lett. {\bf
80}, 2277 (1998).

\bibitem{More}
J.N. Moore and E.P.S. Shellard, hep-ph/9808336.

\bibitem{Jose}
K.D. Olum and J.J. Blanco-Pillado, Phys. Rev. Lett. {\bf 84}, 4288
(2000). 

\bibitem{OS}
K.D. Olum and X. Siemens, Nucl. Phys. {\bf B611}, 125 (2001)

\bibitem{OSV}
K.D. Olum, X. Siemens and A. Vilenkin, Phys. Rev. {\bf D66}, 043501
(2002). 

\bibitem{VOV}
V. Vanchurin, K.D. Olum and A. Vilenkin, Cosmic string scaling in flat
space, gr-qc/0501040.

\bibitem{witten}
E. Witten, Nucl. Phys. {\bf B249}, 557 (1985).

\bibitem{AHDD}
N. Arkani-Hamed, S. Dimopoulos and G. Dvali, Phys. Rev. {\bf D59},
086004 (1999).

\bibitem{Randall}
L. Randall and R. Sundrum, Phys. Rev. Lett. {\bf 83}, 3370 (1999).

\bibitem{DT}
G.R. Dvali and S.-H. Tye, Phys. Lett. {\bf B450}, 72 (1999).

\bibitem{Tye}
S. Sarangi and S.-H. Tye, Phys. Lett. {\bf B536}, 185 (2002).

\bibitem{DV}
G. Dvali and A. Vilenkin, JCAP {\bf 0403}, 010 (2004).

\bibitem{CMP}
E.J. Copeland, R.C. Myers and J. Polchinski, JHEP {\bf 0406}, 013
(2004). 

\bibitem{Polchinski}
J. Polchinski, Introduction to cosmic F- and D-strings,
hep-th/0412244. 

\bibitem{Jeannerot}
R. Jeannerot, J. Rocher and M. sakellariadou, Phys. Rev. {\bf D68},
103514 (2003).

\bibitem{JST}
N.T. Jones, H. Stoica and S.H. Tye, Phys. Lett. {\bf B563}, 6 (2003). 

\bibitem{JJP}
M.G. Jackson, N.T. Jones and J. Polchinski, Collisions of F- and
D-strings, hep-th/0405229.
i
\bibitem{KKLMMT}
S. Kachru, R. Kallosh, A. Linde, J. Maldacena, L. McAllister and
S.P. Trivedi, JCAP {\bf 0310}, 013 (2003).

\bibitem{Damour2}
T. Damour and A. Vilenkin, Phys. Rev. {\bf D71}, 063510 (2005).

\bibitem{Maria}
M. Sakellariadou, JCAP {\bf 0504}, 003 (2005).

\bibitem{SaVi}
M. Sakellariadou and A. Vilenkin, Phys. Rev. {\bf D42}, 349 (1990).

\bibitem{VV}
T. Vachaspati and A. Vilenkin, Phys. Rev. {\bf D35}, 1131 (1987).

\bibitem{SP}
U.-L. Pen and D.N. Spergel, Phys. Rev. {\bf D51}, 4099 (1995).

\bibitem{CS}
E.J. Copeland and P.M. Saffin, hep-th/0505110.

\bibitem{TWW}
S.H. Tye, I. Wasserman and M. Wyman, Phys. Rev. {\bf D71}, 103508
(2005). 

\bibitem{Kaspi} 
V.M. Kaspi, J.H. Taylor and M.F. Ryba, Astrophys. J. {\bf 428}, 713
(1994). 

\bibitem{Levon}
M. Wyman, L. Pogosian and I. Wasserman, Phys. Rev. {\bf D72}, 032513
(2005). 

\bibitem{Fraisse}
A. Fraisse, astro-ph/0503402.

\bibitem{Damour}
T. Damour and A. Vilenkin, Phys. Rev. {\bf D64}, 064008 (2001).

\bibitem{Sazhin}
M. Sazhin {\it et.~al.}, Mon. Not. Roy. Astro. Soc. {\bf 343}, 353
(2003). 

\bibitem{Sazhin2}
M. Sazhin {\it et.~al.}, astro-ph/0506400.

\bibitem{AV86}
A. Vilenkin, Nature {\bf 322}, 613 (1986).

\bibitem{Tye2}
B. Shlaer and S.H. Tye, hep-th/0502242.

\bibitem{Vach}
T. Vachaspati and A. Vilenkin, Phys. Rev. Lett, {\bf 67}, 1057 (1951).

\bibitem{Bohdan}
V. Berezinsky, B. Hnatyk and A. Vilenkin, Phys. Rev. {\bf D64}, 043004
(2001). 

\bibitem{Blasi}
V. Berezinsky, P. Blasi and A. Vilenkin, Phys. Rev. {\bf D58}, 103515
(1998). 


\end{thebibliography}
\end{document}